\documentclass[msmath,amssymb,aps,amsart,12pt]{revtex4-1}
\usepackage[utf8]{inputenc}
\usepackage{amsmath}
\usepackage{graphicx}
\usepackage{multirow}

\begin{document}

\title{Analytical solutions for the Timoshenko beam theory with Free-Free boundary conditions}

\author{R.~A. M\'endez-S\'anchez}
\author{A. A. Fern\'andez-Mar\'in}
\affiliation{Instituto de Ciencias F\'isicas, Universidad Nacional Aut\'onoma de
M\'exico, Apartado Postal 48-3, 62210 Cuernavaca Mor., Mexico}

%\affil{Instituto de Ciencias F\'isicas, Universidad Nacional Aut\'onoma de M\'exico, Apartado Postal 48-3, 62210 Cuernavaca Mor., Mexico}

%\maketitle

\begin{abstract}
Timoshenko's theory for bending vibrations of a beam has been extensively studied since its development nearly one hundred years ago.
Unfortunately there are not many analytical results. 
The results above the critical frequency inclusive haeve been tested only recently.  
Here an analytical expression for the solutions of the Timoshenko equation for free-free boundary conditions, below the critical frequency, is obtained. 
The analytical results are compared with recent experimental results reported of aluminum and brass beams
The agreement is excellent, with an error of less than 3\%, for the aluminum beam and of 5.5\% for the brass beam.
Some exact results are also given for frequencies above the critical frequency.
\end{abstract}

\maketitle


\section*{Introduction}

Bars and beams are very important since they are present in practically all engineered structures and their vibrations characteristics are of fundamental interest.
%Among the four kind of vibrations 
The flexural vibrations of beams, in particular, are still under intense research since they have several applications in engineering. 
For instance the bending vibrations are very important in the design, modelling and test of macro-structures as high buildings, towers and bridges~\cite{SuTangLiu,Yong-Woo,SamadzadRafiee-Dehkharghani}.
In contrast, in microstructures as cantilevers used in atomic force microscopes, and carbon nanotubes, the bending vibrations are also of great interest~\cite{ArashWang,PatraGopalakrishnanGanguli}.
These vibrations are also of interest in high speed rotating equipment~\cite{DeFeliceSorrentino}. 

The flexural vibrations were first described by Bernoulli and Euler resulting in a fourth-order equation~\citep{Graff}. 
Rayleigh modified latter the Bernoulli-Euler theory introducing the angular mass or rotational inertia. 
It was up to the 1920's when Timoshenko developed a theory, known as Timoshenko beam theory (TBT), that includes shear effects~\citep{Timoshenko}.
This theory predicts a critical frequency above which a second spectrum appears. 
The existence of the second spectrum was under intense debate during several decades~\citep{GeistMcLaughlin,OReillyTurcotte,ChanWangSoReid2002,Stephen2006,Bhaskar2009,BhashyamPrathap,Diaz-de-Anda,StephenPuchegger2006,LevinsonCooke,Elishakoff}.
It was only until very recent times~\citep{Diaz-de-AndaEtAl,MonsivaisEtAl,BronsThomsen} that this theory was corroborated experimentally up to frequencies above the critical frequency showing the existence of the critical frequency as well as of the second TBT spectrum.

Although the literature on the Timoshenko beam is vast, there are not many analytical results for the vibrations of beams within this framework. 
In this work, approximate analytical expressions for normal-mode frequencies and amplitudes are given.
This is done for free-free symmetric boundary conditions. 
The obtained results are valid for frequencies below the critical frequency.
In what follows we briefly review the Timoshenko beam theory. 
Then the normal mode frequencies and amplitudes are obtained approximately for free-free (F-F) %and clamped-clamped (C-C) 
boundary conditions (BC).
%The simply supported-simply supported (SS-SS) BC are also given, for completeness. 
The results are compared with those reported in the literature for an aluminum beam, and a brass beam; the agreement is excellent. 
Some analytical results for frequencies above the critical frequency are also given. 
A brief conclusion follows.

\section*{Timoshenko beam theory}
\begin{figure}
	\includegraphics[width=\columnwidth]{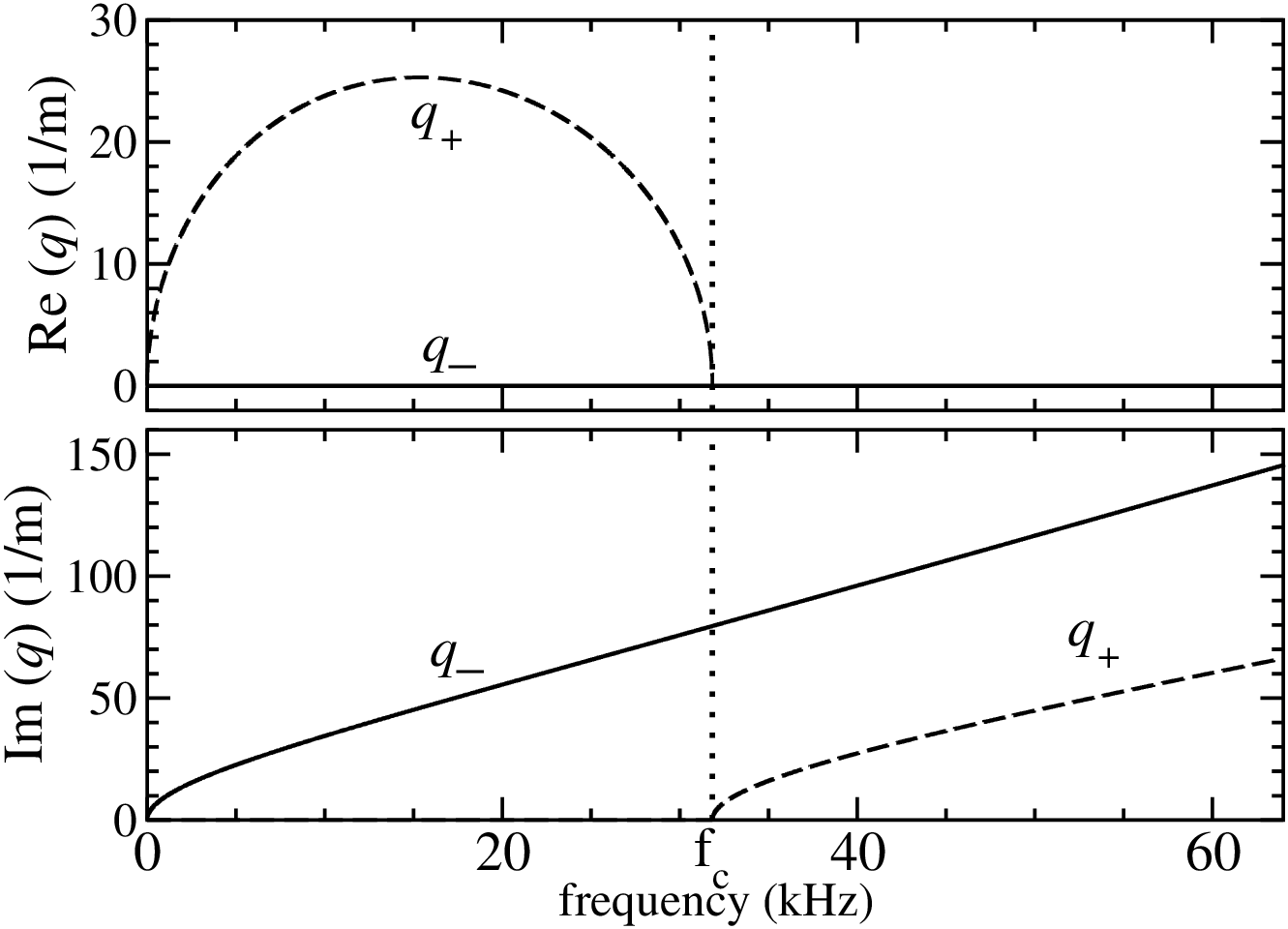}
	\caption{}
	%In the upper and lower panels the real and imaginary parts of the dispersion relation [see Eq.~(\ref{eq:Solcharac})], respectively, are shown for a beam of rectangular cross-section with parameters given in Table~\ref{Tab.ResumResul}}
	\label{fig:qss}
\end{figure}

The transversal displacement $\xi$ and the angular variable $\Psi$ in the Timoshenko beam theory satisfy
\begin{eqnarray}
\kappa GA\left(\frac{\partial\Psi}{\partial z}-\frac{\partial^{2}\xi}{\partial z^{2}}\right)+\rho A\frac{\partial^{2}\xi}{\partial t^{2}} & = & 0\label{Eq:TwoDifferentialEquations1}\\
\kappa GA\left(\frac{\partial\xi}{\partial z}-\Psi\right)+EI\frac{\partial^{2}\Psi}{\partial z^{2}}-\rho I\frac{\partial^{2}\Psi}{\partial t^{2}} & = & 0.\label{Eq:TwoDifferentialEquations2}
\end{eqnarray}
where $E$ is the Young modulus, $G$ the shear modulus, $\rho$ the mass density, $I$ the second moment of area, $A$ the cross-sectional area, and $\kappa$ the Timoshenko shear coefficient~\citep{Cowper,Kaneko,Mendez-SanchezMoralesFlores,Franco-VillafaneMendez-Sanchez}.
The previous equations can be rewritten in terms of the transverse displacement only as
\begin{equation}
\label{eq.1}
\frac{EI}{\rho A}\frac{\partial^4\xi}{\partial z^4}-\frac{I}{A}\Big(1+\frac{E}{\kappa G}\Big)
\frac{\partial^4\xi}{\partial z^2\partial t^2}+\frac{\partial^2\xi}{\partial t^2}+\frac{\rho I}
{\kappa G A}\frac{\partial^4\xi}{\partial t^4}=0.
\end{equation}
This equation can be separated for normal modes. Assuming
\begin{equation}
 \xi(z,t)=e^{\mathrm{i}\omega t}\chi(z)
\label{Eq.separation1}
\end{equation}
one gets
\begin{equation}
\label{eq:chi}
\frac{d^4\chi(z)}{dz^4}-\alpha\frac{d^2\chi(z)}{dz^2}+\frac{\beta}{4}\chi(z)=0,
\end{equation}
where 
\begin{equation}
\alpha=-\frac{\rho\omega^2}{M_\mathrm{r}} \quad \mathrm{and} \quad \beta=\frac{4\rho^2\omega^2}{\kappa GE}(\omega^2-\omega_\mathrm{c}^2), \nonumber
\end{equation}
being $M_\mathrm{r}$ the reduced modulus
\begin{equation}
\frac{1}{M_\mathrm{r}}=\frac{1}{E}+\frac{1}{\kappa G},
\end{equation}
and
\begin{equation}
\omega_\mathrm{c}^2=\frac{\kappa G A}{\rho I}
\end{equation}
the angular critical frequency. The characteristic equation associated to Eq.~(\ref{eq:chi}) is then
\begin{equation}
\label{eq:charac}
q^4-\alpha q^2 +\frac{\beta}{4}=0,
\end{equation} 
whose roots are
\begin{eqnarray}
\label{eq:Solcharac}
   q_\pm^2(\omega) & = & \frac{\alpha\pm\sqrt{\alpha^2-\beta}}{2} \nonumber \\
                   &  = &  -\frac{\rho\omega^2}{2M_\mathrm{r}}\pm
                         \frac{1}{2}\sqrt{\frac{\rho^2\omega^4}{M_\mathrm{r}^2}-\frac{4\rho^2\omega^2}{\kappa GE}(\omega^2-\omega_\mathrm{c}^2)}, \qquad
\end{eqnarray}
that give the dispersion relation. 
Fig.~\ref{fig:qss} shows $q_\pm$ as a function of the angular frequency $\omega$.  
In this figure one can observe that $q_-$ is imaginary in the whole frequency range, whereas $q_+$ changes, from real to imaginary, at the critical frequency $\omega_\mathrm{c}$. 
Thus, for frequencies below the critical frequency, $q_+=k_+$ is real and $q_-=\mathrm{i}k_-$ is imaginary, with $k_-$ real. 
In what follows all equations will be written in terms of $k_+$ and $k_-$, for frequencies below the critical frequency.
The inverse relation 
\begin{equation}
    \label{eq.frequency}
    \omega^2=\frac{\omega_\mathrm{c}^2}{2}
    -\frac{\overline{\mathrm{M}}\, q^2}{\rho}
    \pm \sqrt{\left(\frac{\overline{\mathrm{M}}\, q^2}{\rho}-\frac{\omega_\mathrm{c}^2}{2}\right)^2 -\frac{\kappa G E q^4}{\rho^2}},
\end{equation}
gives the frequency in terms of the wave number with
\begin{equation}
    \overline{\mathrm{M}}=\frac{\kappa G + E}{2},
\end{equation}
the average modulus.

The beam of length $L$ will be symmetrically disposed between $z=-L/2$ and $z=L/2$. Since the boundary conditions will also be symmetric, the solutions of Eq.~(\ref{eq.1}) can be separated in even and odd components as 
\begin{subequations}
	\label{eq:solevenodd}
	\begin{equation}
	\label{eq.7even}
	\chi_n^{(\mathrm{e})}(z) = a_n\cos(k_-z) + b_n\cosh(k_+z),
	\end{equation}
	\begin{equation}
	\label{eq.7odd}
	\chi_n^{(\mathrm{o})}(z) = a'_n\sin(k_-z) + b'_n \sinh(k_+z),
	\end{equation}
\end{subequations}
where the amplitudes $a_n$, $a'_n$, $b_n$, and $b'_n$ are complex constants to be determined by the BC. 

\section*{Solutions for symmetric boundary conditions}

\begin{figure}
	\includegraphics[width=\columnwidth]{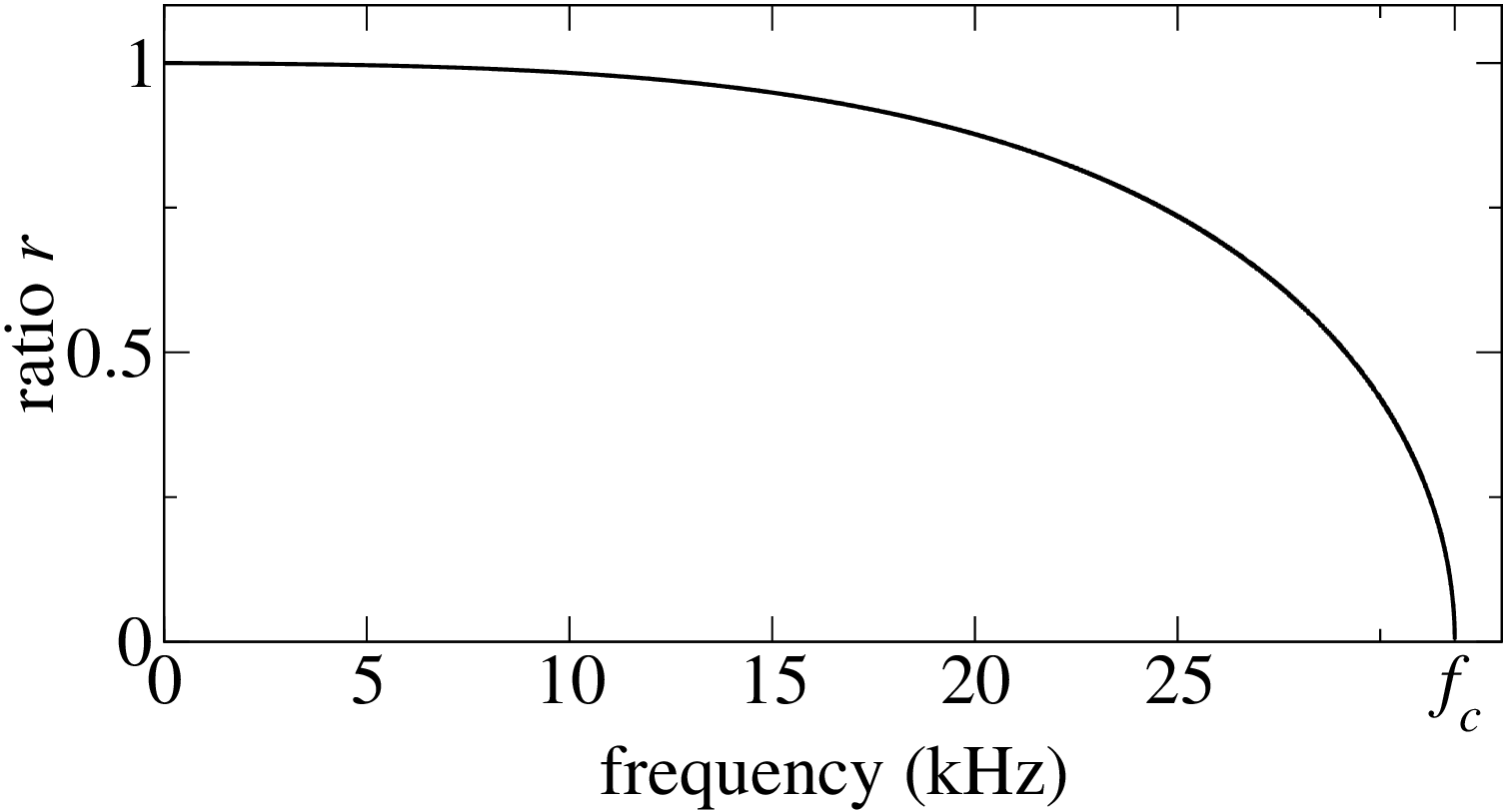}
	\caption{}
	%Fraction $r$ (continuous line) for free-free boundary conditions and $R$ (dashed line) for clamped-clamped boundary conditions, defined in Eqs.~(\ref{Eq.rff}) and~(\ref{Eq.Rcc}), respectively. The same parameters of tables~\ref{Tab.ResumResul} and~\ref{Tab.ResumResulcc} were used.}
	\label{fig:f_pm}
\end{figure}

Following references \citep{GeistMcLaughlin,LevinsonCooke,Diaz-de-AndaEtAl} the Free-Free boundary conditions for the time independent transverse displacement $\chi(z)$ can be obtained as:
%
%\subsection*{Free-Free boundary conditions}
%
%The F-F boundary conditions read
%
\begin{subequations}
	\label{eq:FFbc}
	\begin{equation}
	\left[\frac{d^2 \chi}{dz^2}+\frac{\rho\omega^2}{\kappa G}\chi\right]_{z=\pm L/2}=0,
	\end{equation}    
and
	\begin{equation}
	\left[\frac{d^3\chi}{dz^3}+\frac{\rho\omega^2}{M_\mathrm{r}}\frac{d\chi}{dz}\right]_{z=\pm L/2}=0.
	\end{equation}
\end{subequations}
The substitution of the solution~(\ref{eq:solevenodd}) into the boundary conditions~(\ref{eq:FFbc}) results in the following system equations: %which can be written in matrix form as
%\begin{widetext}
\begin{equation}
	\label{eq:mtx1}
	\left[ 
	\begin{array}{cc}
	\left( \frac{\rho\omega^2}{\kappa G} -k_-^2 \right) \cos\left(\frac{k_-L}{2}\right) & 
	\left( \frac{\rho\omega^2}{\kappa G} + k_+^2 \right) \cosh\left(\frac{k_+L}{2}\right) 
	\\ 
	-\left(\frac{\rho\omega^2}{M_\mathrm{r}} - k_-^2 \right) k_- \sin\left(\frac{k_-L}{2}\right) & 
	\left( \frac{\rho\omega^2}{M_\mathrm{r}} + k_+^2 \right) k_+ \sinh\left(\frac{k_+L}{2}\right)
	\end{array} 
	\right]\!\!
	\left( \begin{array}{c}
	a_n \\ b_n \end{array} \right)\!\! =\! \!
	\left( \begin{array}{c}
	0 \\ 0 \end{array} \right),
\end{equation}
and
\begin{equation}
	\label{eq:mtx2}
	\left[ 
	\begin{array}{cc}
	\left( \frac{\rho\omega^2}{\kappa G} -k_-^2 \right) \sin\left(\frac{k_-L}{2}\right) & 
	\left( \frac{\rho\omega^2}{\kappa G} + k_+^2 \right) \sinh\left(\frac{k_+L}{2}\right) 
	\\ 
	\left(\frac{\rho\omega^2}{M_\mathrm{r}} - k_-^2 \right) k_- \cos\left(\frac{k_-L}{2}\right) & 
	\left( \frac{\rho\omega^2}{M_\mathrm{r}} + k_+^2 \right) k_+ \cosh\left(\frac{k_+L}{2}\right)
	\end{array} 
	\right]\!\!
	\left( \begin{array}{c}
	a_n' \\ b_n' \end{array} \right)\!\! = \!\!
	\left( \begin{array}{c}
	0 \\ 0 \end{array} \right) ,
\end{equation}
%\end{widetext}
%
for the symmetric and antisymmetric cases, respectively.
Non-trivial solutions for these equations are obtained when the determinants $\Delta_\mathrm{e}$ and $\Delta_\mathrm{o}$ of the matrices given in the previous equations vanish, for the even and odd cases, respectively. That is, 
\begin{equation}
\label{eq.trase}
\Delta_\mathrm{e}= r \tan\left(\frac{k_-\mathrm{L}}{2}\right)
+ \tanh\left(\frac{k_+\mathrm{L}}{2}\right)=0, 
\end{equation}
and
\begin{equation}
\label{eq.traso}
\Delta_\mathrm{o}=
r^{-1}\tan\left(\frac{k_-\mathrm{L}}{2}\right)
- \tanh\left(\frac{k_+\mathrm{L}}{2}\right)=0, 
\end{equation}
for even and odd modes of Eq.~(\ref{eq:solevenodd}), respectively. Here 
\begin{equation}
r=\frac{\left(\frac{\rho \omega^2}{M_\mathrm{r}}-k_-^2\right)
\left( \frac{\rho\omega^2}{\kappa G} + k_+^2  \right)
k_- }{\left( \frac{\rho\omega^2}{\kappa G} - k_-^2 \right)
\left(\frac{\rho \omega^2}{M_\mathrm{r}} + k_+^2  \right)k_+ }.
\label{Eq.rff}
\end{equation}
%
%where the upper (lower) sign corresponds to even (odd) modes. 
One can notice in Eqs.~(\ref{eq.trase}) and~~(\ref{eq.traso}) that, below the critical frequency,  $\tanh\left(\frac{k_+\mathrm{L}}{2}\right)\rightarrow 1$ rapidly as the frequency increases. 
Also, the fraction $r$ is close to $1$ for low frequencies (See Fig.~\ref{fig:f_pm}). 
Thus Eqs.~(\ref{eq.trase}) and~(\ref{eq.traso}) can be approximated to
\begin{equation}
\label{eq.Approx1}
\tan(k_-L/2)\approx \mp1, 
\end{equation}
that implies
\begin{equation}
 k_-^{(n)}=\frac{(2n-1)\pi}{2L}, \qquad \mathrm{ with } \qquad n=1,2,\dots.
 \label{Eq.KnFF}
\end{equation}
%
%In what follows $K_n=k_-^{(n)}$ will be used. 
Inserting this result in Eq.~(\ref{eq.frequency}) gives the following analytical expression for the normal-mode frequencies:
\begin{equation}\label{eq:omegan}
\omega_{n}^2= \frac{\omega^2_c}{2}+\frac{\overline{\mathrm{M}} \left(2n-1\right)^2 \pi^2}{ 4 \rho L^2} 
-\sqrt{\left[\frac{\omega^2_c}{2}+\frac{\overline{\mathrm{M}} \left(2n-1\right)^2\pi^2}{4\rho L^2} \right]^2-\frac{\kappa G E \left(2n-1\right)^4\pi^4}{16 \rho^2L^4}}.
\end{equation}
The Euler-Bernoulli result, coming from the last term of the radicand, is found at low values of $n$.
%\begin{widetext}
%	\begin{equation}\label{eq:omegan}
%	\omega_{n}^2=\frac{1}{2}\frac{12DK_n^2}{\rho h^3}\Big(1+\frac{h^3\mu'}{12D}\Big)+\frac{\omega^2_c}{2}
%	\pm\frac{1}{2}\sqrt{\Big[\frac{12DK_n^2}{\rho h^3}\Big(1+\frac{h^3\mu'}{12D}\Big)+\omega^2_c\Big]^2-\frac{4D\omega_\mathrm{c}^2}{\rho h}K_n^4}.
%	\end{equation}
%\end{widetext}
%
%It is remarkable that, within this approximation, the normal-mode frequencies depend only on $k_-$ but not on $k_+$. 
%Once $\omega_n$ is given, $k_+(\omega_n)$ can be obtained from Eq.~(\ref{eq:Solcharac}).

\begin{figure}
	\includegraphics[width=\columnwidth]{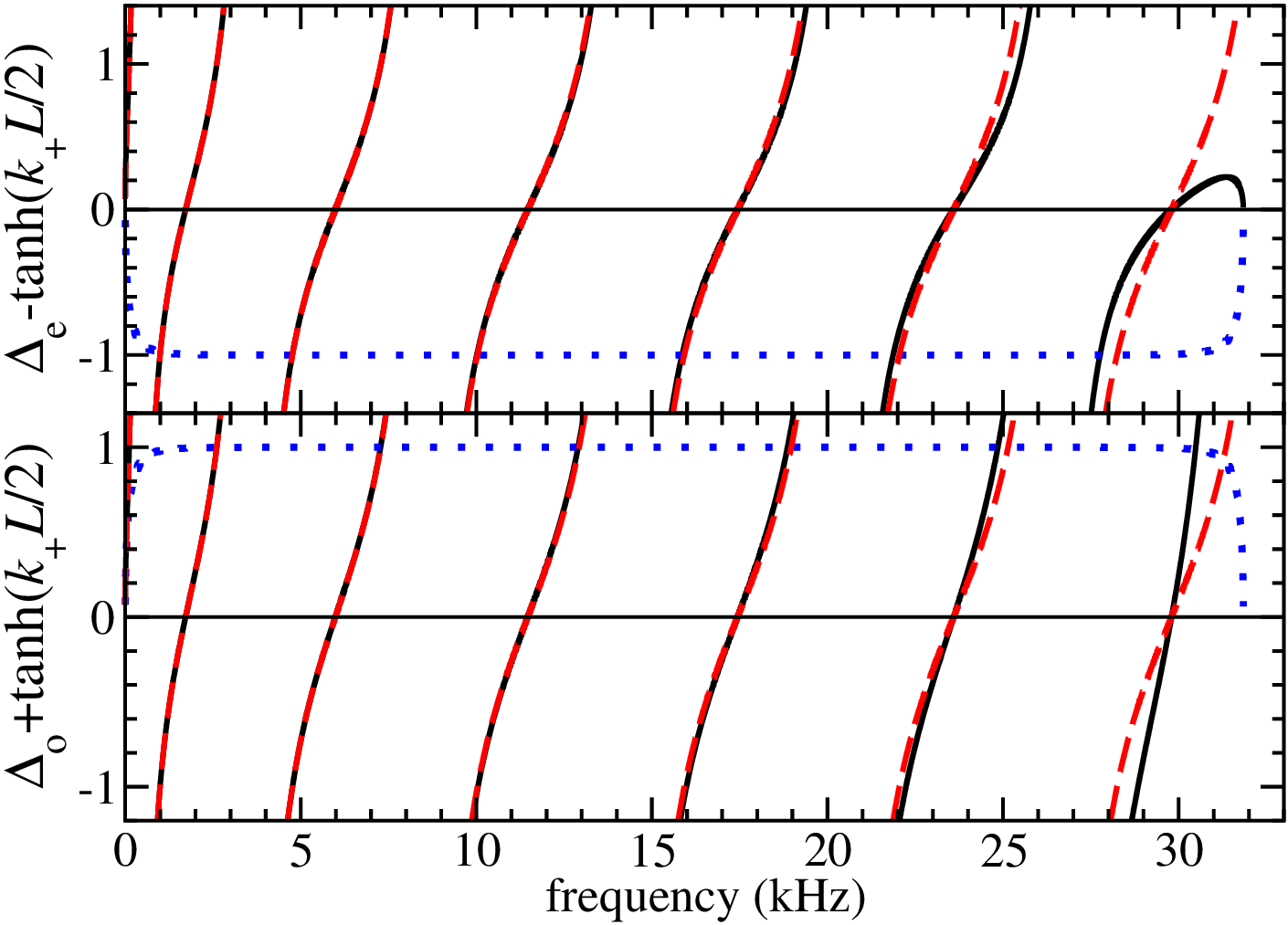}
	\caption{}
	%(Color online) Graphical solution of Eqs.~(\ref{eq.trase}) and~(\ref{eq.traso}) (intersection between the dotted (black) and the black continuous curves). In the upper and lower panels the even and odd cases are given, respectively. 	The dashed (red) curve is $\tan(k_-L/2)$. 	The  parameters used here are those given in Table~\ref{Tab.ResumResul}.}
	\label{fig:trasc}
\end{figure}

The beam of Ref.~\citep{Diaz-de-AndaEtAl} will be studied now.
Fig.~\ref{fig:trasc} shows the graphical solution of the transcendental Eqs.~(\ref{eq.trase}) and~(\ref{eq.traso}) which occurs at the intersection between the (blue) dotted and the (black) continuous curves for the even (a) and odd (b) cases. 
In the same figure a dashed (red) curve, which corresponds to $\tan(k_-L/2)$, is also plotted. The approximate solutions are obtained when the dashed curve is $\pm 1$. 
In Table~\ref{Tab.ResumResul} we compare the analytical predictions given by Eq.~(\ref{eq:omegan}) with those of Ref.~\citep{Diaz-de-AndaEtAl} and with the numerical solution of Eqs.~(\ref{eq.trase}) and~(\ref{eq.traso}). 
The approximate solutions have an excellent agreement with the numerical and experimental results of Ref.~\citep{Diaz-de-AndaEtAl}; the percentage error is less than 3\%.
%The aluminum beam of rectangular cross-section has dimension: $L=0.5$ m, height $a=0.0252$~m and width $b=0.0504$~m, and the elastic constants are $G=26.92$~GPa, $E=67.42$~GPa and $\rho=2699.04$~kg/m$^{3}$.

%
\begin{table}[!hbt]
	%\centering
	\caption{Normal mode frequencies for the aluminum beam of rectangular cross-section with $L=0.5$ m, height $a=0.0252$~m and width $b=0.0504$~m with free ends. The elastic constants are $G=26.92$~GPa, $E=67.42$~GPa and $\rho=2699.04$~kg/m$^{3}$.}
	%\begin{center}
	\begin{tabular}{llll}
		\hline
		Mode & Experiment of & TBT$\qquad$  & Eq.~(\ref{eq:omegan})  \\
	      number & Ref.~\citep{Diaz-de-AndaEtAl} (kHz) & (kHz) & (kHz) \\
		\hline
		1      & 1.0211  & 1.00062 & 0.99212 ($2.84$\%)  \\
		2      & 2.6594  & 2.60559 & 2.60685  ($2.32$\%)  \\
		3      & 4.8462  & 4.75908 & 4.76048  ($1.77$\%)  \\
		4      & 7.3878  & 7.27210 & 7.27716  ($1.50$\%)  \\
		5      & 10.163  & 10.0153 & 10.0282  ($1.33$\%)  \\
		6      & 13.082  & 12.8989 & 12.9272  ($1.18$\%)  \\
		7      & 16.081  & 15.8629 & 15.9179  ($1.01$\%)  \\
		8      & 19.1335 & 18.8651 & 18.9642  ($0.88$\%)  \\
		9      & 22.1699 & 21.8731 & 22.0426  ($0.57$\%)  \\
		10     & 25.1638 & 24.8548 & 25.1379  ($0.10$\%)  \\
		11     & 28.0593 & 27.7656 & 28.2404  ($0.65$\%)  \\
		12     & 30.6109 & 30.4859 & 31.3437  ($2.40$\%)  \\
		$f_{\mathrm{c}}$ &     & 31.850      &             \\
		\hline
	\end{tabular}
	\label{Tab.ResumResul}
	%\end{center}
\end{table}

Eq.~(\ref{eq:omegan}) will now be compared with those of the brass beam of Ref.~\citep{BronsThomsen}. In Table~\ref{Tab.NewResults} the normal-mode frequencies obtained from Eq.~\ref{eq:omegan} are given. The experimental results of Ref.~\citep{BronsThomsen} are also given. As it can be seen, the results present an error of less than {\bf 5.5\% }. Thus an excellent agreement is also obtained with the results of Ref.~\citep{BronsThomsen}.

\begin{table}[!hbt]
	%\centering
	\caption{Normal mode frequencies for the brass beam of rectangular cross-section with $L=0.6865$ m, height $a=0.060$~m and width $b=0.00402$~m with free ends. The elastic constants are $G=34.8$~GPa, $E=91.27$~GPa and $\rho=8427$~kg/m$^{3}$.}
	%\begin{center}
	\begin{tabular}{llll}
		\hline
		Mode & Experiment of & TBT$\qquad$  & Eq.~(\ref{eq:omegan})  \\
	      number & Ref.~\citep{BronsThomsen} (kHz) & (kHz) & (kHz) \\
		\hline
		1      & 0.433  & 0.419  &  0.416   ($4.08 $\%)  \\
		2      & 1.116  & 1.103  &  1.103   ($1.18 $\%)  \\
		3      & 2.106  & 2.039  &  2.037   ($3.39 $\%)  \\
		4      & 3.225  & 3.150  &  3.148   ($2.44 $\%)  \\
		5      & 4.409  & 4.381  &  4.379   ($0.69 $\%)  \\
		6      & 5.677  & 5.690  &  5.690   ($-0.23 $\%)  \\
		7      & 6.968  & 7.049  &  7.054   ($-1.22 $\%)  \\
		8      & 8.320  & 8.438  &  8.452   ($-1.56 $\%)  \\
		9      & 9.674  & 9.842  &  9.871   ($-2.00 $\%)  \\
		10     & 10.993 & 11.249 &  11.303  ($-2.74 $\%)  \\
		11     & 12.308 & 12.650 &  12.742  ($-3.40 $\%)  \\
		12     & 13.588 & 14.033 &  14.184  ($-4.20 $\%)  \\
		13     & 14.787 & 15.375 &  15.626  ($-5.37 $\%)  \\		
		14     & 15.577 & 16.609 &  17.067  ($-8.73 $\%)  \\				$f_{\mathrm{c}}$& & 17.04      &             \\
		\hline
	\end{tabular}
	\label{Tab.NewResults}
	%\end{center}
\end{table}

Although this work only gives results for the case of $\omega<\omega_\mathrm{c}$, Eqs.~(\ref{eq.trase}) and~(\ref{eq.traso}) can be used for frequencies larger than the critical frequency. In this case the hyperbolic tangent results in a trigonometric tangent. Changing $k_+$ by $\mathrm{i}Q_+$, with $Q_+$ real, one gets
\begin{equation}
\left[\frac{\left(\frac{\rho \omega^2}{M_\mathrm{r}}-k_-^2\right)
\left( \frac{\rho\omega^2}{\kappa G} - Q_+^2  \right)
k_- }{\left( \frac{\rho\omega^2}{\kappa G} - k_-^2 \right)
\left(\frac{\rho \omega^2}{M_\mathrm{r}} - Q_+^2  \right)Q_+ }
\right]^{\pm 1}
\tan\left(\frac{k_-\mathrm{L}}{2}\right)-\tan\left(\frac{Q_+\mathrm{L}}{2}\right)=0,
\end{equation}
where the upper (lower) sign corresponds to even (odd) modes.

\section*{Wave amplitudes}
The shape of the modes can be calculated directly once the wave numbers and the normal-mode frequencies are obtained. Using the first of Eqs.~(\ref{eq:mtx1}) in Eqs.~(\ref{eq:solevenodd}), one gets
\begin{subequations}
\begin{eqnarray}
\chi_n^{(\mathrm{e})}(z) &=& A_n
\left( \frac{\rho\omega_n^2}{\kappa G} + k_{+,n}^2\right) 
\cosh\left( \frac{k_{+,n}L}{2} \right)
\cos\left(K_n z\right) \nonumber \\
&-& A_n \left( \frac{\rho\omega_n^2}{\kappa G} - K_n^2\right)
\cos\left(\frac{K_nL}{2}\right)
\cosh\left(k_{+,n} z\right) ,
\label{eq:wfns}
\end{eqnarray}
and
\begin{eqnarray}
\chi_n^{(\mathrm{o})}(y) &=& A'_n
\left( \frac{\rho\omega_n^2}{\kappa G} + k_{+,n}^2\right)
\sinh\left( \frac{k_{+,n}L}{2} \right)
\sin\left(K_n z\right) \nonumber \\
&-& A'_n \left( \frac{\rho\omega_n^2}{\kappa G} - K_n^2\right) 
\sin\left(\frac{K_nL}{2}\right)
\sinh\left(k_{+,n} z\right) ,
\label{eq:wfna}
\end{eqnarray}
\end{subequations}
for the even and odd cases, respectively. $A_n$ and $A'_n$ are normalization constants. 
The wave amplitudes can also be calculated exactly for $f>f_\mathrm{c}$ as before changing $k_{+,n}$ by $\mathrm{i} Q_+$. In all cases the hyperbolic functions are converted in trigonometric functions with $Q_+$ instead of $k_{+,n}$ in the arguments.

\begin{figure}
	\includegraphics[width=\columnwidth]{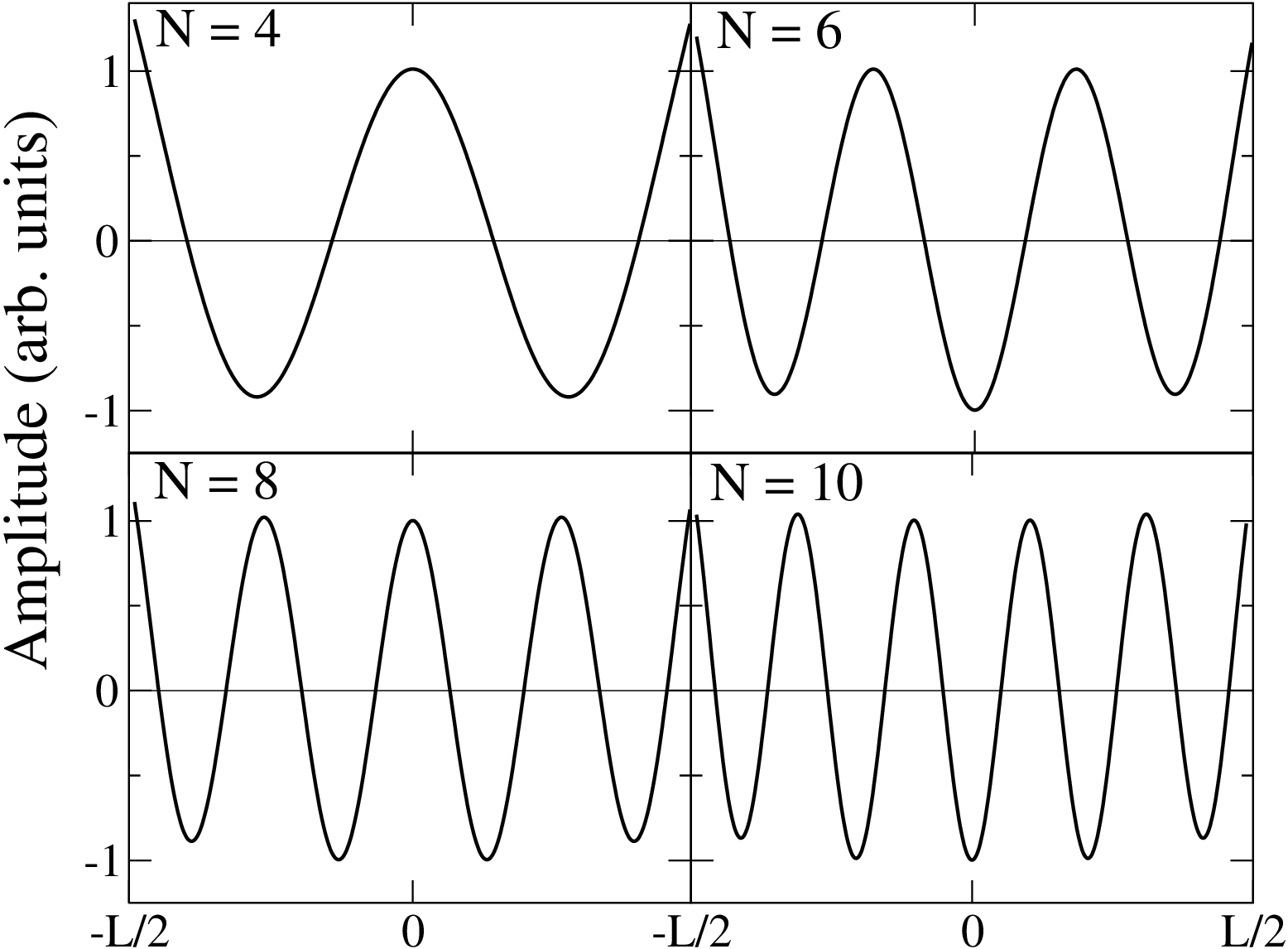}
	\caption{}
	%Wave amplitudes for $N=4,6,8$ and 10 nodes for a rectangular beam with: free ends (upper row), clamped ends (middle row) and simply supported ends (lower row).}
	\label{fig:waves}
\end{figure}

In Fig.~\ref{fig:waves} the wave amplitudes for $N= 4,6,8$ and $10$ nodes for a rectangular beam with free ends are given. %It is noticeable in this figure that the wave amplitudes of the SS-SS and C-C cases are quite similar one to each other, with small differences at the borders.
 
\section*{Conclusions}

In this work we obtained approximate expressions for the normal-mode frequencies of the Timoshenko beam theory below the critical frequency. This was done for a beam with free-free boundary conditions.
The obtained normal-mode frequencies show an excellent agreement with published experimental results, inclusive up to frequencies close to the critical frequency.
Some analytical results were also given for frequencies above the critical frequency.

\section*{Acknowledgments}
We thank G. B\'aez and M. Mart\'{\i}nez for useful comments. 
A.A.F.-M. acknowledges a postdoctoral fellowship from DGAPA-UNAM. This work was supported by DGAPA-UNAM and by CONACYT under projects PAPIIT IN111021 and CB-2016/284096, respectively.
The authors acknowledge the kind hospitality of Centro Internacional de Ciencias A.C. for group meetings celebrated frequently there. 

%======================================================================

\newpage

\newpage

\newpage

\newpage

%\begin{figure}
%	\includegraphics[width=\columnwidth]{Fig4.eps}
%	\caption{}
	%(Color online) Graphical solution of Eq.~(\ref{eq.trascc}) given by the intersection between the continuous curve and the dotted one. In the upper and lower panels are given the even and odd cases, respectively. The dashed (red) curve is $\tan(k_-L/2)$. The same parameters of the F-F case, given in Table~\ref{Tab.ResumResul}, are used in this case.}
%	\label{fig:trasccc}
%\end{figure}

\newpage

\newpage

\vspace*{5cm}\centerline{\bf }

\centerline{\bf Figure captions}

\vspace*{1cm}\noindent Figure 1: In the upper and lower panels the real and imaginary parts of the dispersion relation [see Eq.~(\ref{eq:Solcharac})], respectively, are shown for a beam of rectangular cross-section with parameters given in Table~\ref{Tab.ResumResul}\\

\noindent Figure 2: Fraction $r$ for free-free boundary conditions defined in Eqs.~(\ref{Eq.rff}). The same parameters of table~\ref{Tab.ResumResul} were used.\\

\noindent Figure 3: (Color online) Graphical solution of Eqs.~(\ref{eq.trase}) and~(\ref{eq.traso}) (intersection between the dotted (black) and the black continuous curves). In the upper and lower panels the even and odd cases are given, respectively. The dashed (red) curve is $\tan(k_-L/2)$.	The  parameters used here are those given in Table~\ref{Tab.ResumResul}.\\

%\noindent Figure4: (Color online) Graphical solution of Eq.~(\ref{eq.trascc}) given by the intersection between the continuous curve and the dotted one. In the upper and lower panels are given the even and odd cases, respectively. The dashed (red) curve is $\tan(k_-L/2)$. The same parameters of the F-F case, given in Table~\ref{Tab.ResumResul}, are used in this case.\\

\noindent Figure 4: Wave amplitudes for $N=4,6,8$ and $10$ nodes for a rectangular beam with free ends.
\\

\nocite{*}
\bibliography{biblioTBT.bib}

\providecommand*{\mcitethebibliography}{\thebibliography}
\csname @ifundefined\endcsname{endmcitethebibliography}
{\let\endmcitethebibliography\endthebibliography}{}
\begin{mcitethebibliography}{25}
\providecommand*{\natexlab}[1]{#1}
\providecommand*{\mciteSetBstSublistMode}[1]{}
\providecommand*{\mciteSetBstMaxWidthForm}[2]{}
\providecommand*{\mciteBstWouldAddEndPuncttrue}
  {\def\EndOfBibitem{\unskip.}}
\providecommand*{\mciteBstWouldAddEndPunctfalse}
  {\let\EndOfBibitem\relax}
\providecommand*{\mciteSetBstMidEndSepPunct}[3]{}
\providecommand*{\mciteSetBstSublistLabelBeginEnd}[3]{}
\providecommand*{\EndOfBibitem}{}
\mciteSetBstSublistMode{f}
\mciteSetBstMaxWidthForm{subitem}
{(\emph{\alph{mcitesubitemcount}})}
\mciteSetBstSublistLabelBeginEnd{\mcitemaxwidthsubitemform\space}
{\relax}{\relax}

\bibitem[Su \emph{et~al.}(2016)Su, Tang, and Liu]{SuTangLiu}
R.~Su, T.~Tang and K.~Liu, \emph{Engineering Structures}, 2016, \textbf{120},
  116--132\relax
\mciteBstWouldAddEndPuncttrue
\mciteSetBstMidEndSepPunct{\mcitedefaultmidpunct}
{\mcitedefaultendpunct}{\mcitedefaultseppunct}\relax
\EndOfBibitem
\bibitem[Kim()]{Yong-Woo}
Y.-W. Kim, \emph{Structural Engineering and Mechanics}, \textbf{62},
  247--258\relax
\mciteBstWouldAddEndPuncttrue
\mciteSetBstMidEndSepPunct{\mcitedefaultmidpunct}
{\mcitedefaultendpunct}{\mcitedefaultseppunct}\relax
\EndOfBibitem
\bibitem[Samadzad and Rafiee-Dehkharghani(2020)]{SamadzadRafiee-Dehkharghani}
M.~Samadzad and R.~Rafiee-Dehkharghani, \emph{International Journal of Dynamics
  and Control}, 2020, \textbf{8}, 459--476\relax
\mciteBstWouldAddEndPuncttrue
\mciteSetBstMidEndSepPunct{\mcitedefaultmidpunct}
{\mcitedefaultendpunct}{\mcitedefaultseppunct}\relax
\EndOfBibitem
\bibitem[Arash and Wang(2012)]{ArashWang}
B.~Arash and Q.~Wang, \emph{Computational Materials Science}, 2012,
  \textbf{51}, 303--313\relax
\mciteBstWouldAddEndPuncttrue
\mciteSetBstMidEndSepPunct{\mcitedefaultmidpunct}
{\mcitedefaultendpunct}{\mcitedefaultseppunct}\relax
\EndOfBibitem
\bibitem[Patra \emph{et~al.}(2020)Patra, Gopalakrishnan, and
  Ganguli]{PatraGopalakrishnanGanguli}
A.~K. Patra, S.~Gopalakrishnan and R.~Ganguli, \emph{Acta Mechanica}, 2020,
  \textbf{231}, 1159--1171\relax
\mciteBstWouldAddEndPuncttrue
\mciteSetBstMidEndSepPunct{\mcitedefaultmidpunct}
{\mcitedefaultendpunct}{\mcitedefaultseppunct}\relax
\EndOfBibitem
\bibitem[De~Felice and Sorrentino(2019)]{DeFeliceSorrentino}
A.~De~Felice and S.~Sorrentino, \emph{Meccanica}, 2019, \textbf{54},
  1029--1055\relax
\mciteBstWouldAddEndPuncttrue
\mciteSetBstMidEndSepPunct{\mcitedefaultmidpunct}
{\mcitedefaultendpunct}{\mcitedefaultseppunct}\relax
\EndOfBibitem
\bibitem[Graff(1991)]{Graff}
K.~Graff, \emph{Wave Motion in Elastic Solids}, Dover Publications, 1991\relax
\mciteBstWouldAddEndPuncttrue
\mciteSetBstMidEndSepPunct{\mcitedefaultmidpunct}
{\mcitedefaultendpunct}{\mcitedefaultseppunct}\relax
\EndOfBibitem
\bibitem[Timoshenko(1921)]{Timoshenko}
P.~S. Timoshenko, \emph{The London, Edinburgh, and Dublin Philosophical
  Magazine and Journal of Science}, 1921, \textbf{41}, 744--746\relax
\mciteBstWouldAddEndPuncttrue
\mciteSetBstMidEndSepPunct{\mcitedefaultmidpunct}
{\mcitedefaultendpunct}{\mcitedefaultseppunct}\relax
\EndOfBibitem
\bibitem[Geist and McLaughlin(1997)]{GeistMcLaughlin}
B.~Geist and J.~McLaughlin, \emph{Applied Mathematics Letters}, 1997,
  \textbf{10}, 129--134\relax
\mciteBstWouldAddEndPuncttrue
\mciteSetBstMidEndSepPunct{\mcitedefaultmidpunct}
{\mcitedefaultendpunct}{\mcitedefaultseppunct}\relax
\EndOfBibitem
\bibitem[ORe(1996)]{OReillyTurcotte}
\emph{Journal of Sound and Vibration}, 1996, \textbf{198}, 517--521\relax
\mciteBstWouldAddEndPuncttrue
\mciteSetBstMidEndSepPunct{\mcitedefaultmidpunct}
{\mcitedefaultendpunct}{\mcitedefaultseppunct}\relax
\EndOfBibitem
\bibitem[Chan \emph{et~al.}(2002)Chan, Wang, So, and Reid]{ChanWangSoReid2002}
K.~T. Chan, X.~Q. Wang, R.~M.~C. So and S.~R. Reid, \emph{Proceedings of the
  Royal Society of London. Series A: Mathematical, Physical and Engineering
  Sciences}, 2002, \textbf{458}, 83--108\relax
\mciteBstWouldAddEndPuncttrue
\mciteSetBstMidEndSepPunct{\mcitedefaultmidpunct}
{\mcitedefaultendpunct}{\mcitedefaultseppunct}\relax
\EndOfBibitem
\bibitem[Stephen(2006)]{Stephen2006}
N.~Stephen, \emph{Journal of Sound and Vibration}, 2006, \textbf{292},
  372--389\relax
\mciteBstWouldAddEndPuncttrue
\mciteSetBstMidEndSepPunct{\mcitedefaultmidpunct}
{\mcitedefaultendpunct}{\mcitedefaultseppunct}\relax
\EndOfBibitem
\bibitem[Bhaskar(2009)]{Bhaskar2009}
A.~Bhaskar, \emph{Proceedings of the Royal Society A: Mathematical, Physical
  and Engineering Sciences}, 2009, \textbf{465}, 239--255\relax
\mciteBstWouldAddEndPuncttrue
\mciteSetBstMidEndSepPunct{\mcitedefaultmidpunct}
{\mcitedefaultendpunct}{\mcitedefaultseppunct}\relax
\EndOfBibitem
\bibitem[Bhashyam and Prathap(1981)]{BhashyamPrathap}
G.~Bhashyam and G.~Prathap, \emph{Journal of Sound and Vibration}, 1981,
  \textbf{76}, 407--420\relax
\mciteBstWouldAddEndPuncttrue
\mciteSetBstMidEndSepPunct{\mcitedefaultmidpunct}
{\mcitedefaultendpunct}{\mcitedefaultseppunct}\relax
\EndOfBibitem
\bibitem[D\'{\i}az-de Anda \emph{et~al.}(2005)D\'{\i}az-de Anda, Pimentel,
  Flores, Morales, Gutiérrez, and M\'endez-Sánchez]{Diaz-de-Anda}
A.~D\'{\i}az-de Anda, A.~Pimentel, J.~Flores, A.~Morales, L.~Gutiérrez and
  R.~A. M\'endez-Sánchez, \emph{The Journal of the Acoustical Society of
  America}, 2005, \textbf{117}, 2814--2819\relax
\mciteBstWouldAddEndPuncttrue
\mciteSetBstMidEndSepPunct{\mcitedefaultmidpunct}
{\mcitedefaultendpunct}{\mcitedefaultseppunct}\relax
\EndOfBibitem
\bibitem[Stephen and Puchegger(2006)]{StephenPuchegger2006}
N.~Stephen and S.~Puchegger, \emph{Journal of Sound and Vibration}, 2006,
  \textbf{297}, 1082--1087\relax
\mciteBstWouldAddEndPuncttrue
\mciteSetBstMidEndSepPunct{\mcitedefaultmidpunct}
{\mcitedefaultendpunct}{\mcitedefaultseppunct}\relax
\EndOfBibitem
\bibitem[Levinson and Cooke(1982)]{LevinsonCooke}
M.~Levinson and D.~Cooke, \emph{Journal of Sound and Vibration}, 1982,
  \textbf{84}, 319--326\relax
\mciteBstWouldAddEndPuncttrue
\mciteSetBstMidEndSepPunct{\mcitedefaultmidpunct}
{\mcitedefaultendpunct}{\mcitedefaultseppunct}\relax
\EndOfBibitem
\bibitem[Elishakoff \emph{et~al.}(2017)Elishakoff, Hache, and
  Challamel]{Elishakoff}
I.~Elishakoff, F.~Hache and N.~Challamel, \emph{International Journal of Solids
  and Structures}, 2017, \textbf{109}, 143--151\relax
\mciteBstWouldAddEndPuncttrue
\mciteSetBstMidEndSepPunct{\mcitedefaultmidpunct}
{\mcitedefaultendpunct}{\mcitedefaultseppunct}\relax
\EndOfBibitem
\bibitem[de~Anda \emph{et~al.}(2012)de~Anda, Flores, Gutiérrez,
  Méndez-Sánchez, Monsivais, and Morales]{Diaz-de-AndaEtAl}
A.~D. de~Anda, J.~Flores, L.~Gutiérrez, R.~Méndez-Sánchez, G.~Monsivais and
  A.~Morales, \emph{Journal of Sound and Vibration}, 2012, \textbf{331},
  5732--5744\relax
\mciteBstWouldAddEndPuncttrue
\mciteSetBstMidEndSepPunct{\mcitedefaultmidpunct}
{\mcitedefaultendpunct}{\mcitedefaultseppunct}\relax
\EndOfBibitem
\bibitem[Monsivais \emph{et~al.}(2016)Monsivais, de~Anda, Flores, Gutiérrez,
  and Morales]{MonsivaisEtAl}
G.~Monsivais, A.~D. de~Anda, J.~Flores, L.~Gutiérrez and A.~Morales,
  \emph{Journal of Sound and Vibration}, 2016, \textbf{375}, 187--199\relax
\mciteBstWouldAddEndPuncttrue
\mciteSetBstMidEndSepPunct{\mcitedefaultmidpunct}
{\mcitedefaultendpunct}{\mcitedefaultseppunct}\relax
\EndOfBibitem
\bibitem[Brøns and Thomsen(2019)]{BronsThomsen}
M.~Brøns and J.~J. Thomsen, \emph{Journal of Sound and Vibration}, 2019,
  \textbf{459}, 114856\relax
\mciteBstWouldAddEndPuncttrue
\mciteSetBstMidEndSepPunct{\mcitedefaultmidpunct}
{\mcitedefaultendpunct}{\mcitedefaultseppunct}\relax
\EndOfBibitem
\bibitem[Cowper(1966)]{Cowper}
G.~R. Cowper, \emph{Journal of Applied Mechanics}, 1966, \textbf{33},
  335--340\relax
\mciteBstWouldAddEndPuncttrue
\mciteSetBstMidEndSepPunct{\mcitedefaultmidpunct}
{\mcitedefaultendpunct}{\mcitedefaultseppunct}\relax
\EndOfBibitem
\bibitem[Kaneko(1975)]{Kaneko}
T.~Kaneko, \emph{Journal of Physics D: Applied Physics}, 1975, \textbf{8},
  1927--1936\relax
\mciteBstWouldAddEndPuncttrue
\mciteSetBstMidEndSepPunct{\mcitedefaultmidpunct}
{\mcitedefaultendpunct}{\mcitedefaultseppunct}\relax
\EndOfBibitem
\bibitem[Méndez-Sánchez \emph{et~al.}(2005)Méndez-Sánchez, Morales, and
  Flores]{Mendez-SanchezMoralesFlores}
R.~Méndez-Sánchez, A.~Morales and J.~Flores, \emph{Journal of Sound and
  Vibration}, 2005, \textbf{279}, 508--512\relax
\mciteBstWouldAddEndPuncttrue
\mciteSetBstMidEndSepPunct{\mcitedefaultmidpunct}
{\mcitedefaultendpunct}{\mcitedefaultseppunct}\relax
\EndOfBibitem
\bibitem[Franco-Villafa\~ne and
  Méndez-Sánchez(2016)]{Franco-VillafaneMendez-Sanchez}
J.~A. Franco-Villafa\~ne and R.~A. Méndez-Sánchez, \emph{Journal of
  Mechanics}, 2016, \textbf{32}, 515--518\relax
\mciteBstWouldAddEndPuncttrue
\mciteSetBstMidEndSepPunct{\mcitedefaultmidpunct}
{\mcitedefaultendpunct}{\mcitedefaultseppunct}\relax
\EndOfBibitem
\end{mcitethebibliography}

\end{document}